\documentclass[cmp]{svjour}
\usepackage{amsmath,amsfonts, amssymb}

\begin{document}
\title{The von Neumann entropy and information rate for ideal quantum
Gibbs ensembles}

\author{Oliver Johnson \and Yuri Suhov}
\institute{Statistical Laboratory, 
DPMMS/CMS, University of Cambridge, Cambridge CB3 0WB, UK
\email{otj1000@cam.ac.uk, yms@statslab.cam.ac.uk} Fax: +44 1223 337956} 
\date{Received: date / Accepted: date}
\titlerunning{von Neumann entropy and information rate}
%
\communicated{name}
\maketitle
\begin{abstract}
A  model of a  quantum information  source is  proposed, based  on the
Gibbs ensemble  of ideal (free) particles (bosons or  fermions).  
We identify the (thermodynamic) von Neumann entropy as the information
rate and establish
the classical Lempel--Ziv universal coding algorithm in Grassberger's
form for such a source. This generalises the Schumacher theorem to the 
case of non-IID qubits.
\end{abstract}

\newcommand{\ov}[1]{\overline{#1}}
\newcommand{\ep}{{\mathbb {E}}}
\newcommand{\pr}{{\mathbb {P}}}
\newcommand{\Z}{{\mathbb {Z}}}
\newcommand{\R}{{\mathbb {R}}}
\newcommand{\K}{{\mathbb {K}}}
\newcommand{\var}{\rm Var\;}
\newcommand{\rar}[1]{{#1}^L}
\newcommand{\rart}[1]{\overline{#1}^L}
\newcommand{\nm}[1]{\| #1 \|}
\newcommand{\rev}[1]{#1^{\rm{rev}}}
\newcommand{\bx}[2]{B_{\vec{#1}}(#2)}
\newcommand{\ku}[2]{\vec{k}_{#1}(#2)}
\newtheorem{assumption}[theorem]{Assumption}

\newenvironment{pfof}[2]{\removelastskip\vspace{6pt}\noindent
 {\it Proof  #1.}~\rm#2}{\par\vspace{6pt}}

\section {Introduction and basic facts}
In classical information theory, the fundamental unit is a `bit', and the 
model behind it is a random variable taking values 0 and 1 with probability
1/2. We often refer to a sequence of random variables as a source -- note
that the physics of the way in which the random variables are generated is
irrelevant, and that results on data-compression rely only on the statistics
of long `strings'.

In the newer quantum information theory, the fundamental unit is a `qubit',
which is associated with a two-dimensioanal complex Hilbert space.
Here the structure   is much richer, since states can be not only
$|0\rangle$ or $|1\rangle$ but any complex linear combination in between. 
However, the definition of a general quantum source producing a 
sequence of qubits remains open.

So far, the theory of quantum 
data compression has confined itself to the case of 
qubits  emitted by an IID (independent identically distributed)  source. 
Here, a qubit is a general
$2 \times 2$ density matrix $\sigma$, and the assumption
of independence is that the state of $n$ qubits is described by the tensor
power $\sigma^{\otimes n}$. 
IID qubits can be implemented as photon pulses emitted by a laser. However,
this model does not allow natural
entanglement, and hence lacks interesting physical
properties. Even the most enthusiastic 
proponents of modern quantum information theory
consider the IID assumption as ``an unfortunate 
restriction'' (see Nielsen and Chuang \cite{nielsen}, p. 554). 
It was noted that attempts to reliably produce a qubit 
string by using various 
``random processes, such as the preparation and detection of
photon pairs ... or atoms in thermal beam... suffer from inescapable
signal degradation, ... as the probability of randomly generating
the appropriate conditions decreases exponentially'' (\cite{sackett}, p. 256).

On the other hand, in practice, qubits can be modelled by using physical
particles or spins -- electrons or atoms.
Recent experimental results in quantum entanglement
(see Sackett et al \cite{sackett}) indicate that perhaps the most reliable 
way to prepare a string of quantum qubits is to couple 
quantum particles in a coherent way. In the experiment 
reported in \cite{sackett}, these were ions of $^9$Be$^+$ 
interacting, approximately,  
via a Dicke-Lamb type potential and arranged in a one-dimensional lattice.
A similar approach was put forward in \cite{jkp}. 
Most recent experiments with physical implementation of Shor's quantum
factorisation algorithm also use quantum particle systems as a material base of
a computational device \cite{lieven}.
For a mathematician, this stimulates 
interest in rigorous analysis of information coding 
methods for sources represented by {\it ensembles} of 
quantum particles. 

The first step in this direction would be to consider the eigenvector
distribution of a Gibbs density matrix of a large system of quantum 
particles or `spins'. An eigenvector $\phi$ of the density matrix 
can in principle be identified as a result of a quantum `measurement'
and the probability that in the grand canonical Gibbs ensemble the
system chooses a pure eigenstate $|\phi\rangle\langle\phi |$ is 
proportional to $\exp \left( - \beta \mu n -\beta\lambda \right)$. 
Here $n$ is the number of particles in state 
$|\phi\rangle\langle\phi |$,
$\mu$ represents the chemical potential (and $z=e^{\beta\mu}$
the `fugacity'), $\lambda$ is the corresponding
eigenvalue and $\beta =1/\kappa T$ where the $T$ is the absolute 
temperature
and $\kappa$ the Boltzmann constant. The idea of our approach is
that the corresponding eigenvector may usually be 
represented as a long sequence of numbers (`digits'). If 
the quantum ensemble carries `enough randomness', such a sequence
can be treated as a sample of a random process or field.  It
seems interesting to analyse such a process or field from the
point of view of (classical) information theory.

A natural (and simplest) example to consider is a system
of free quantum particles in a volume $\Lambda\subset \R^d$ 
(an open bounded domain with piecewise smooth boundary 
$\partial\Lambda$). 
The interaction here is manifested through the chosen statistics 
(Bose or Fermi). The grand canonical Gibbs ensemble in $\Lambda$ 
is described by a quasi-free bosonic or fermionic  
density matrix $\rho^{\Lambda}_\pm$ in the Fock Hilbert space 
${\cal F}^{\Lambda}_\pm$
associated with volume $\Lambda$ (index $\pm$ indicates the
Bose or Fermi statistics). Such a state
is generated by the one-particle Hamiltonian
$H$ $(=H_1^\Lambda )$, a self-adjoint operator in the
one-particle complex Hilbert space ${\cal H}$ $(= 
{\cal H}_1^\Lambda )$, given
values of the thermodynamical parameters $\beta$ and $\mu$. 

A typical model is where ${\cal H}_1^\Lambda = L_2(\Lambda )$ and 
operator $H$ is minus one-half of the Laplacian 
with a `classical' boundary condition
on $\partial\Lambda$, see for example \cite{br2} Sections 5.2.4 and 5.2.5.
In this case we assume that a) $\beta >0$ and b)  
$\mu >0$ for bosons and $-\infty <\mu <\infty$ for fermions. 
A lattice version of such a model is where 
${\cal H}^\Lambda$ is the Hilbert space whose 
(complex) dimension equals $\#\,(\Lambda\cap \Z^d)$,
the number of points $\vec{l}=(l_1,\ldots ,l_d)\in \Z^d$ 
with integer components $l_j$ within $\Lambda$. Here,   
$H$ may be minus one-half of the discrete
Laplacian, again with a `classical' boundary condition
on $\partial\left(\Lambda\cap \Z^d\right)$. 

Suppose that $H^\Lambda $ has a pure discrete spectrum
and the eigenvalues of $H_1^\Lambda $ (counted with their
multiplicities) are 
$\gamma^\Lambda_{\vec{n}}$, with $\min_{ \vec{n} \in {\cal N}} 
\gamma^\Lambda_{\vec{n}} = 0$.
Here $\vec{n}$ runs over a finite or denumerable 
set ${\cal N}$ $(={\cal N}^\Lambda )$ 
and $\sum_{\vec{n}\in{\cal N}}\exp\;(-\beta\gamma^\Lambda_{\vec{n}})<\infty$
for all $\beta>0$. For instance, if $\Lambda\subset{\R}^d$ 
is a cube $(-L/2,L/2)^d$ 
and $H = -1/2\Delta$ with periodic 
boundary conditions, ${\cal N}$ coincides with the 
integer cubic lattice
$ \Z^d$ and $\gamma^\Lambda_{\vec{n}} = 
4\pi^2|\vec{n}|^2/L^2$ where $|\vec{n}| = \left(n_1^2+\ldots 
+n_d^2\right)^{1/2}$, $\vec{n} \in  \Z^d$.  
 
An eigenvector 
$\phi$ of the quasi-free density matrix $\rho^{\Lambda}$ is 
associated with a sequence of occupation 
numbers $\vec{k} = \{k_{\vec{n}}, \vec{n} \in {\cal N}\}$ (we will also  
write $\phi =\phi^\Lambda_{\vec{k}}$). More precisely, $k_{\vec{n}}$ 
is a non-negative integer equal to the number of particles
in the eigenstate of $H$ with the eigenvalue $\gamma_{\vec{n}}^{\Lambda}$; 
in the fermion case, $k_{\vec{n}}=0$ or $1$. It is convenient to 
set ${\K}_+= \Z_+:=\{0,1,2,\ldots \}$ for the 
boson and ${\K}_-=\{0,1\}$ for the fermion case.
In both cases, the number of non-zero entries 
$k_{\vec{n}}$ in a given $\vec{k}$ is finite, with the sum 
$\sum_{\vec{n} \in{\cal N}} k_{\vec{n}}$ representing 
the number of particles. The corresponding 
eigenvalue is
\begin{equation} \label{eq:lambda}
\lambda (=\lambda^\Lambda_{\vec{k}})=\exp\left(-\beta
\sum_{\vec{n} \in {\cal N}} k_{\vec{n}}
(\mu + \gamma^{\Lambda}_{\vec{n}} )\right). \end{equation} 
Thus the probability 
that the system will be found in pure state 
$\left| \phi^{\Lambda}_{\vec{k}}\rangle
\langle\phi^{\Lambda}_{\vec{k}} \right|$ 
is proportional to  
\begin{equation}
\exp\;\left( -\beta \sum_{ \vec{n} \in {\cal N}}
k_{\vec{n}} (\mu + \gamma^{\Lambda}_{\vec{n}}) \right)
=\prod_{\vec{n}\in{\cal N}} 
\exp \left( -\beta (\gamma_{\vec{n}}^{\Lambda} +\mu ) k_{\vec{n}} \right). 
\label{eq:probdist} \end{equation}
In other words, a free quantum ensemble produces an `array' 
$\vec{K} = \{K_{\vec{n}}, \vec{n} \in{\cal N} \}$ of random variables
$K_{\vec{n}}$ with probability determined by Equation (\ref{eq:probdist}). 
Throughout we use the convention that upper case letters refer to random
variables, and lower case letters to the values that they take.
This product form means that random variables
$K_{\vec{n}}$, $\vec{n}\in{\cal N}$, are independent (but not identically
distributed). The marginal distribution of $K_{\vec{n}}$
is geometric for bosons and two-point for fermions.
Let ${\cal P}$ $(={\cal P}^\Lambda_\pm)$ denote the 
induced probability distribution on ${\cal K}_\pm = 
{\K}_\pm^{\cal N}$, supported by the set ${\cal K}_\pm^0$ of arrays
with finitely many non-zero components); it is convenient to think 
that ${\cal P}$ is determined by the quadruple $\left({\cal H}^\Lambda,
H^\Lambda,\beta,\mu\right)$. 

Now assume that $\{\Lambda\}$ is  an increasing sequence of volumes in
${\R}^d$   eventually  covering   the  whole   of   ${\R}^d$,  writing
$\Lambda\nearrow{\R}^d$. In  is convenient to think  that $\Lambda$ is
the result of  the homothetic dilation of a  fixed open bounded domain
$\Lambda^0\subset{\R}^d$ containing  the origin and  with a piece-wise
smooth boundary $\partial\Lambda^0$ consisting of finitely many smooth
parts. In  the above example,  we can think  of $\Lambda^0$ as  a unit
cube $(-1/2,1/2)^d$  and $\Lambda = (-L/2,L/2)^d$  as its dilation
by the linear factor $L$.

A question arises then: what are the properties of the `source' 
$({\cal K},{\cal P}^\Lambda )$? To what extent can classical coding 
theory be applied to such a source (or rather a sequence of 
sources, as $\Lambda\nearrow{\R}^d$)? 
Some classical results are easily extended to 
the the case of $({\cal K},{\cal P}^\Lambda )$ (after all,
${\cal P}^\Lambda $ is a product-distribution, albeit
not stationary). For example, an asymptotic 
equipartition property (AEP) for $({\cal K},{\cal P}^\Lambda)$
is fairly straightforward (see Proposition \ref{prop:lln}). 
[This property can
be considered as an analog of the famous Shannon--McMillan--Breiman
Theorem in the situation under consideration.]
The corresponding information rate coincides with 
the von Neumann entropy per unit `volume' of the 
limiting quantum free ensemble. 

However, results beyond the AEP, such as the classical Lempel--Ziv universal 
encoding algorithm, are more tricky to establish. The Lempel-Ziv
algorithm, in its various forms, is perhaps the most popular
encoding method in modern information transmission. 
The idea of the algorithm, in the form of `parsing'
originally proposed by Ziv and Lempel \cite{lempel} is very simple.
Suppose we have a sequence $x_0,x_1,x_2,\ldots$ of `letters'
from an `alphabet' (say, $x_i\in\{0,1\}$ (the binary alphabet)).
We put a marker sign (say, a semi-colon ;) after $x_0$. If
$x_1\not =x_0$, we put the marker sign after $x_1$, otherwise
(i.e., if $x_0=x_1$), we put it after $x_2$. Continuing this 
procedure, given that the last marker sign was after $x_j$,
we put the next marker sign after letter $x_{j'}$, $j'>j$, if 
for all $s=1,\ldots ,j'-j-1$, the
`word' $(x_{j+1},\ldots ,x_{j+s})$ is among the `blocks' formed
between the subsequent marker signs already in place, but
the `word' $(x_{j+1},\ldots ,x_{j'})$ has not been seen before.

This gives rise to the following encoding method:
each new parsed word has a `header' (the word less the 
last letter) which has been seen before. Thus, to `encode'
this bit of sequence $x_0,x_1,x_2,\ldots$ we need only to
indicate the place where the header was seen in the past
and in addition encode the last letter of the new block.

The popularity of this algorithm is due to its universal character (no 
knowledge of the properties of the source is required to implement it),
and to the fact that asymptotically it achieves the data compression limit.
However, this convergence is slow, leading to adaptions of the algorithm,
including
the so-called Grassberger \cite{grassberger} form of the algorithm which 
also suits the multi-dimensional situation ($d >1$).

In Sections \ref{sec:prelim} and \ref{sec:lznot} 
we state our main results (see Theorem
\ref{thm:main}), that the Lempel--Ziv algorithm is valid (again with
the von Neumann entropy as the information rate). In the higher-dimensional
case, we establish this result in Grassberger's form, and in the 
one-dimensional case, we also prove it in terms of the classical Lempel--Ziv
algorithm. The proofs are given 
in Sections \ref{sec:lowbd} -- \ref{sec:lz}.

Our assumption on quadruple $({\cal H},H,
\mu,\beta )$ follow the basic model outlined above
where $\Lambda = (-L/2,L/2)^d$, ${\cal H}^\Lambda = 
L_2(\Lambda )$ and $H^\Lambda = -1/2\Delta$ with 
periodic boundary conditions. [The lattice version of this
model can also be easily incorporated]. We consider
fixed $\beta >0$ and $\mu >0$ for bosons and $-\infty < \mu < \infty$ for 
fermions. 
Although this formally excludes the Bose--Einstein condensation,
the fact is that the condensation is largely irrelevant to 
our results. We intend to discuss this issue in a separate paper.
Furthermore, many of the properties obtained in this paper  
can be in turn extended to systems with
interaction. The corresponding results are now in preparation.

We would like to point out an essential non-uniqueness of the definition of
the quantum entropy (or entropies), see \cite{connes}. From this point of
view, it would be interesting to clarify the relation of various concepts of
quantum entropy with quantum information theory.
\section{Preliminary results} \label{sec:prelim}
Our main assumption is that a) the set ${\cal N}$ 
coincides with $ \Z^d$, the cubic lattice,
and so the collection of `arrays' ${\cal K}_\pm^0 \subset {\K}_\pm^{ \Z^d}$ 
consists of functions on $ \Z^d$ with finite 
supports and with values in ${\K}_+ = \Z_+$ for bosons and 
${\K}_- = \{0,1\}$ for fermions,
b) the eigenvalues $\gamma^{\Lambda}_{\vec{n}}$, $\vec{n} \in  \Z^d$,
of the one-particle Hamiltonian $H^\Lambda$
are of the form $\theta (\nm{\vec{n}}/L)$, where c) $L =L(\Lambda )$
is a parameter increasing to $\infty$ as sequence of volumes
$\Lambda\nearrow \Z^d$ (it will be covenient to assume that
$L$ simply runs over the set of natural numbers), and d) 
$\theta$: $[0, \infty) \to
[0,\infty )$ is a given continuous function, such that $\theta(x) >0$ for
$x>0$, and such that the following integral is finite:
\begin{equation} \label{eq:vonneu}
\int_{{\R}^d} \left( \mp \log\,
\left(1 \mp e^{-\beta (\theta (\nm{\vec{y}})+\mu )}
\right) + \beta\left(\theta (\nm{\vec{y}})+\mu\right)
\left( \frac{e^{-\beta (\theta (\nm{\vec{y}})+\mu )}}{1 \mp
e^{-\beta (\theta (\nm{\vec{y}})+\mu )}} \right) \right) d\vec{y},
\end{equation}
where we take the first choice of all the $\mp$ for bosons, for all 
$\beta,\mu >0$,  and the second choice for for fermions, for all $\beta >0$,
$-\infty < \mu < \infty$.

Parameter $L$ can be though of as a `linear size'
of $\Lambda$ and henceforth is used instead of $\Lambda$.
In other words, we fix a sequence of positive
numbers $L\to\infty$ replacing $\Lambda\nearrow{\R}^d$, say
$L = 1,2, \ldots$.
As was suggested, it is convenient to think that
$\Lambda$ is the cube $(-L/2,L/2)^d$. 

In the model where $H^\Lambda = - \Delta/2$ with
periodic boundary conditions, $\theta (t)=4\pi^2 t^2$.
\begin{definition} \label{def:vonneu}
Integrals (\ref{eq:vonneu}) are called the 
von Neumann entropy per unit volume in the free boson/fermion
limiting Gibbs ensemble and denoted by $h_\pm$. The restriction
of the integral to a domain $\Gamma\subset{\R}^d$
is denoted by $h_\pm^{\Gamma}$. \end{definition}
\begin{remark} \label{rem:moti} 
The reasoning behind this definition is as 
follows. The probability
measure ${\cal P}^L$ has been specified by Equation (\ref{eq:lambda}) 
as the
product $\times_{\vec{n} \in \Z^d} \pi_{\vec{n}}$ where $\pi_{\vec{n}}$ is 
the  geometric distribution with parameter
$e^{-\beta (\gamma_{\vec{n}}+\mu )}$ for bosons
and the two-point distribution, with $\pi_{\vec{n}} (\{0\}) =  
\frac{1}{1+e^{-\beta (\gamma_{\vec{n}}+\mu )}}$,
$\pi_n (\{1\}) = 
\frac{e^{-\beta (\gamma_{\vec{n}}+ \mu )}}{1+e^{-\beta (\gamma_{\vec{n}}+\mu 
)}}$,
for fermions. The entropy of ${\cal P}^L$ divided by $L^d$
(the volume of $\Lambda$) is simply a Riemann sum for
the integral $h_\pm$ and converges to $h_\pm$ as $L\to\infty$. 
On the other hand, the entropy of ${\cal P}^L$ is equal to
the von Neumann entropy ${\rm{tr}}_{{\cal F}^L_\pm}\rho^L
\log\,\rho^L$ of the density matrix $\rho^L$ corresponding to
the Gibbs ensemble of free quantum particles in $\Lambda$,
for given $\beta$ and $\mu$. \end{remark}
It is easy to check the following law of large numbers.
\begin{proposition} \label{prop:lln}
Consider the random variable $\xi_\pm^L$: 
$\vec{k}\to (1/L^d)\log\,\lambda^L (\vec{k})$, 
$\vec{k}\in{\cal K}_\pm^0$, where $\lambda^L (\vec{k})$
is the eigenvalue of Gibbs ensemble density matrix
$\rho_\pm^L$ determined by function $\vec{k}$: $ \Z\to{\K}_\pm$.
Then, for all $\varepsilon >0$, $\displaystyle{\lim_{L\to\infty}
{\cal P}_\pm^L
\left( \left| \xi^L -h_\pm \right|
\geq \varepsilon\right)} = 0$. 
Also, $\displaystyle{\lim_{L\to\infty}\xi^L(\vec{K}^L)=h_\pm}$ almost surely 
(a.s.) with respect to the product measure ${\cal P}^{\times L}$ on
the Cartesian power ${\cal K}^{\times L}$, with the sequence 
$(\vec{K}^L)$ of ${\cal P}^L$-random elements. \end{proposition}
A straightforward consequence of Proposition \ref{prop:lln} is
\begin{corollary} \label{cor:aep}
List the eigenvalues $\lambda^L (\vec{k})$,
$\vec{k}\in{\cal K}_\pm^0$, in decreasing 
order: $\lambda_{(0)}\geq\lambda_{(1)}\geq\ldots $. Given $\epsilon\in (0,1)$,
select the eigenvalues in their order until the sum of the 
selected $\lambda$'s becomes greater than or equal to
the value $1-\epsilon$ for the first time. Let $M_\pm^L$
denote the number of selected eigenvalues. Then
$\displaystyle{\lim_{L\to\infty}\frac{1}{L^d}\log\;M_\pm^L}
= h_\pm$. \end{corollary}
Definition \ref{def:vonneu}, Remark \ref{rem:moti}, Proposition 
\ref{prop:lln} and Corollary \ref{cor:aep} specify an 
asymptotic equipartition property for probability measures
${\cal P}^L$, and $h_\pm$ can be considered as
an analog of the information rate for $({\cal K}^0,{\cal P}^L)$. 
\section{Main result}\label{sec:lznot}
For the rest of the paper, $\vec{k}\in{\cal K}^0_\pm\;$ is 
a function $ \Z^d\to{\K}_\pm$ with 
compact support; we identify it with the collection
of values $k_{\vec{n}}$, $\vec{n} \in \Z^d$.
Given a probability measure ${\cal P}^L$ on ${\cal K}_\pm = 
{\K}_\pm^{ \Z^d}$, 
$\vec{K}$ stands for an array of random 
variables $\{K_n\}$ representing the random element 
of ${\cal K}_\pm$. When considering the product-measure 
${\cal P}^{\times L}$ on the Cartesian product ${\cal K}^{\times L}$,
we denote by $\vec{K}^L$ the ${\cal P}^L$-random element
of ${\cal K}$.
\begin{definition}
Given $\vec{k}\in{\cal K}$, 
$\vec{u} = (u_1,\ldots, u_d)\in \Z^d$ and $s \geq 1$, 
define the cubic box $\bx{u}{s}$ to have  
bottom corner $\vec{u}$ and side $s$:
$$\bx{u}{s} = \{ \vec{n} = (n_1,\ldots ,n_d) \in \Z^d:  
u_i \leq n_i \leq u_i+s-1, \mbox{ for all
$i = 1, \ldots d$} \}.$$
Write $\vec{k}_{\vec{u}}(s) = \{ k_{\vec{n}}$: $\vec{n} \in \bx{u}{s} \}$ 
for the set of values of $\vec{k}$ confined to this box.

Now define $\rar{r}_{\vec{u}}(\vec{k})$ 
to be the size of the smallest box with bottom corner at position $\vec{u}$
with values different to all the others with bottom corner 
in $\bx{1}{L}$, $\vec{1} = (1,\ldots, 1) \in \Z^d$:
$$\rar{r}_{\vec{u}}(\vec{k})= 
\inf \{s \geq 1: \vec{k}_{\vec{u}}(s) \neq \vec{k}_{\vec{v}}(s) 
\mbox{ for all $\vec{v} \neq \vec{u} \in \bx{1}{L}$} \}.
$$
For $\vec{K}$ a random array, we define the random variable 
$\rar{R}_{\vec{u}}(\vec{K})$ in the same way. \end{definition}
These $\rar{R}_{\vec{u}}$ have been studied by authors such as 
Grassberger \cite{grassberger}, Kontoyiannis and Suhov \cite{kontoyiannis},
Quas \cite{quas} and Shields \cite{shields}, \cite{shields2} 
first in the one-dimensional
case and later for higher dimensions, partly because they
serve as good entropy estimators for an ergodic process
with a suitable degree of mixing. For example, 
Theorem 1 of \cite{quas} shows: 

{\it If the array $ \vec{K}$
is  generated by a $ \Z^d$-invariant ergodic 
probability measure on ${\cal K}_\pm$ with entropy 
$h$, under a Doeblin condition,}  
$$\lim_{L\to\infty}
\sum_{\vec{u} \in \bx{1}{L}} \frac{\rar{R}_{\vec{u}}(\vec{K})}{L^d\log L} 
= \frac{1}{h},\;
\lim_{L\to\infty}\sum_{\vec{u} \in \bx{1}{L}} \frac{\log L}
{L^d \rar{R}_{\vec{u}}(\vec{K})}=h ,\;
\mbox{a.s.}$$
Our Theorem \ref{thm:main} below shows how a similar result looks
for sequences $({\cal K}^{\times L}_\pm,{\cal P}^{\times L}_\pm)$: 
\begin{theorem} \label{thm:main}
For all fixed $\zeta >0$, on ${\cal K}_\pm^{\times L}$, 
$$\lim_{L\to\infty} \sum_{\vec{u} \in \bx{1}{\zeta L}} 
\frac{\log L}{\left(\zeta L\rar{R}_{\vec{u}}(\vec{K}) \right)^d} =
h_\pm^{\bx{0}{\zeta}}, \mbox{  ${\cal P}_\pm^{\times L}$-a.s.} $$
Here $h_\pm^{\bx{0}{\zeta}}$ is the `truncated'
von Neumann entropy (cf Definition \ref{def:vonneu}),
where $\bx{0}{\zeta}$ is the cube $[0,\zeta]^n$.\end{theorem}
We can deal with the case of $\zeta$ increasing with $L$, under
extra assumptions on the behaviour of $\theta$.
\begin{assumption}  \label{ass:theta}
For all $\eta$, there exist $C, \delta$ such that
uniformly in $x > \eta$ for $y < \delta$:
$\theta(x+y)/\theta(x) \leq C$. 
\end{assumption}
\begin{assumption} \label{ass:zeta}
Our $\zeta \rightarrow \infty$, slowly enough that $\zeta/\log L \rightarrow 
0$.
\end{assumption}
\begin{theorem}
If $\theta$ satisfies Assumption \ref{ass:theta} and 
$\zeta$ satisfies Assumption \ref{ass:zeta} then on ${\cal K}_\pm^{\times L}$, 
$$\lim_{L\to\infty} \sum_{\vec{u} \in \bx{1}{\zeta_L L}} 
\frac{\log L}{\left(\zeta_L L\rar{R}_{\vec{u}}(\vec{K}) \right)^d} =
h_\pm, \mbox{  ${\cal P}_\pm^{\times L}$-a.s.} $$
\end{theorem}
\begin{remark}
Alternatively, in the spirit of previous analysis, 
we can average the 
$\rar{R}_{\vec{u}}$ themselves. However, Theorem \ref{thm:main} in our view 
gives a more useful  
result for von Neumann entropy estimation. \end{remark}
 
For the sake of clarity, we focus on the case $\zeta =1$ (though we indicate
in due course
how the case of $\zeta$ varying with $L$ can 
naturally be dealt with) and first prove the one-dimensional 
($d=1$) version of the result for 
geometric variables (bosons), in Sections 
\ref{sec:lowbd} and \ref{sec:uppbd}. In Section \ref{sec:fermdir}, we 
indicate the adaptations needed in the case of two-valued 
variables (fermions), and
in Section \ref{sec:highdim}, we show how the method adapts to the
case of higher dimensions.
We split the proof of the result into 3 parts, corresponding to the 
Lemmas 6, 7 and 8 used in \cite{quas}.
In each case, writing $\rar{E}_{\vec{u}}$ for the entropy of $X_{\vec{u}}$ under
${\cal P}^\Lambda$, we will
show that $\log L/(\rar{R}_{\vec{u}})^d$ is close to $\rar{E}_{\vec{u}}$.
\begin{lemma} \label{lem:mainlow}
For any $\epsilon >0$, then ${\cal P}^{\times L}$-a.s.:
$$ \lim_{L\to\infty}
\frac{1}{L^d} \# \left\{ {\vec{u}}: \rar{R}_{\vec{u}}(\rar{\vec{K}})  
\leq \left( \frac{{\log L}(1-\epsilon)}{\rar{E}_{\vec{u}}}  \right)^{1/d}
 \right\} = 0.$$ \end{lemma}
\begin{lemma} \label{lem:mainupp}
For any $\epsilon >0$, then ${\cal P}^{\times L}$-a.s.:
$$ \lim_{L\to\infty}
\frac{1}{L^d} \# \left\{
{\vec{u}}: \rar{R}_{\vec{u}} (\rar{\vec{K}}) \geq 
\left(\frac{{\log L}(1+\epsilon)}{\rar{E}_{\vec{u}}} \right)^{1/d} \right\}
= 0.$$ \end{lemma}
\begin{lemma} \label{lem:unifupp} 
There exists a constant $c = c(\theta)$ such that
${\cal P}^{\times L}$-a.s.:
$$ \limsup_{L\to\infty} \left( \max_{i \in \bx{1}{R}}
\frac{\rar{R}_{\vec{u}}(\rar{\vec{K}})}{({\log L})^{1/d}} \right) \leq c.$$
\end{lemma}
\section{Proof of lower bound} \label{sec:lowbd}
Recall, in the next two sections, we concentrate on 
the one-dimensional geometric case and consider $\zeta =1$. 
So, an array $\vec{k}\in{\cal K}$ is now a 
`string' $\{k_i$, $i\in \Z\}$, 
where $k_i$ is a non-negative integer. Write
$\vec{k}_i(s)$ for a finite
piece $(k_i, \ldots k_{i+s-1})$ of string $\vec{k}$ of length
$s$ starting at position $i$ 
where $i,s\in \Z$, $s\geq 1$.
Then $\rar{r}_{i}(\vec{k})$ is the length  
of the shortest piece starting at position $i$
with values different to all the others pieces starting 
in $\{1,\ldots ,L\}$: $\rar{r}_{i}(\vec{k})= 
\inf \{s>0: \vec{k}_i(s) \neq \vec{k}_j(s)\mbox{   
for all $j \neq i$ in $\{1,\ldots ,L\}$ } \}$. Accordingly, 
$\rar{r}_{i}(\vec{k})$ is often called the match length.
Here $\theta$ is a continuous function: $[0, \infty) \rightarrow [0, \infty)$ 
and $E_i^L$ is the entropy of the geometric distribution
with parameter $e^{-\beta (\theta (|i|/L)+\mu)}$. 
Write $\theta^*$ for $\sup_{x \in [0,1]} 
\theta (x)$.

We use the idea of a `typical set', familiar from
Ergodic and Information Theory. The aim is to 
show that usually we belong in this typical set
$S$, which provides extra conditions so that the match length
$\rar{R}_i$ cannot be too low too often.
\begin{definition} \label{def:loglik}
For a string $\vec{k}=\{k_i$, $i\in \Z\}$, 
we define the centred log-likelihood:
$$\rar{y}_i(\vec{k}) = - \log {\cal P}^L(K_i = k_i) - \rar{E}_u
= \beta (\mu + \theta (i/L)) (k_i - \ep^L \rar{K}_i),$$
and for the random string $\vec{K}$, we
define $\rar{Y}_i$ for the random variable $\rar{y}_i(\vec{K})$.
Here and below, $\ep^L$ stands for the expectation
relative to ${\cal P}^L$. Define
the typical set by 
$$ \rar{S}_{j,M} = \left\{ \vec{k}: \left| 
\sum_{i=j}^{j+M-1} \rar{y}_i(\vec{k}) \right|
\leq M \theta(j/L) \epsilon' \right\},$$
where $\epsilon' = \beta \epsilon (-\log\,(1-e^{-\theta^*})/2\theta^*$.
\end{definition}

\begin{pfof}{of Lemma \ref{lem:mainlow}}
Now for any sequence $M(i)$ and any $\eta >0$, we deal with the first
$\eta L$ variables separately:
\begin{eqnarray*}
\lefteqn{
\left\{ \vec{k}: \frac{1}{L} \# \left\{  i: \rar{R}_i(\vec{k}) \leq 
\frac{{\log L}(1-\epsilon)}{\rar{E}_i} \right\} > 3 \eta \right\} } \\
& \subseteq &
\left\{ \vec{k}: \# \left\{ i > \eta L : \rar{R}_i(\vec{k}) \leq \frac{\log 
L(1-\epsilon)}{\rar{E}_i}, \vec{k} \in \rar{S}_{i,M(i)} \right\} > 
\eta L \right\} \\
&  & \bigcup  \left\{ \vec{k}: \# \left\{  i > \eta L :
\vec{k} \notin \rar{S}_{i,M(i)} \right\} > \eta L\right\} 
\end{eqnarray*}
We bound the size of the first set in Lemma \ref{lem:propbd}
and the size of the second in Lemma \ref{lem:vdec}.
\qed \end{pfof}
\begin{lemma} \label{lem:propbd}
Given $\eta, \epsilon >0$, we can find a 
sequence $M(i)$ and constant
$C_1(\eta,\epsilon)$ such that
for any $L \geq C_1$ and for any $\vec{k} \in {\cal K}^0$ 
$$  \# \left\{  i  > \eta L : \rar{R}_i(\vec{k}) \leq \frac{\log 
L(1-\epsilon)}{\rar{E}_i}, \vec{k} \in \rar{S}_{i,M(i)} \right\} \leq\eta L.$$
\end{lemma}
\begin{proof}
We can find intervals $J_i$ in which our variables have their means close 
together. Note that $f(x) = 1/(e^{\beta (\mu + x)} -1)$ has derivative
bounded below on $x > e > 0$. Hence, 
given $\theta$ and $\epsilon$, we can calculate 
$N = N(\epsilon)$ and $u_1, \ldots u_N$ with $u_1 = \eta$, $u_N = 1$ such 
that
$$ \frac{1}{e^{\beta (\mu + \theta(u_i))}-1} \leq \frac{1}{e^{\beta (\mu + 
\theta(u_{i+1})}-1} 
+ \epsilon', \mbox{ for $i = 1, \ldots N-1$,} $$
where $\epsilon'$ is from Definition \ref{def:loglik}.
Defining
$J_i = \{ m: m/L \in (u_i, u_{i+1}) \}$, for each $j \in J_i$ 
define $ M(j) = {\log L}(1-\epsilon)/ u_{i+1}.$ 

We compare $\rar{S}_{j,M}$ with
$D_{\gamma,M}$, a set which we can count and control more easily.
For each $\gamma, M$, define $E_\gamma$ for the entropy of 
a geometric distribution with parameter $e^{-\gamma}$ and
$$D_{\gamma,M} = \left\{ \vec{x}_1(M)=(x_1,\ldots ,x_M): 
\sum_{i=1}^M x_i \leq M \left( 
\frac{1}{e^\gamma-1} + \frac{\epsilon E_{\gamma}}{\gamma} \right) \right\}.$$
For $\vec{x}_1(M) \in D_{\gamma,M}$, writing $\pr_\gamma$ for
product measure for independent geometric random variables with
parameter $e^{-\gamma}$:
$$ \pr_{\gamma}(\vec{x}_1(M)) = \exp
\left(  M \log (1- e^{-\gamma}) - \gamma 
\sum_{i=1}^M x_i \right) \geq 
\exp(-M E_\gamma (1 + \epsilon)).$$ 

If $\vec{k} \in \rar{S}_{j,M}$, where $j \in J_i$, taking $\gamma =
\sup_{x \in J_i} \beta (\mu + \theta(x))$:
$$ \sum_{l=j}^{j+M-1} k_l \leq
M \left( \frac{1}{e^{\beta (\mu + \theta(j/L)} -1}  + \epsilon' \right) \leq
M \left( \frac{1}{e^\gamma -1}  + \frac{\epsilon
E_{\gamma}}{\gamma} \right),$$
so $\vec{k}_j(M) = (k_j,\ldots k_{j+M-1}) \in D_{\gamma,M}$.
We therefore know 
that if $\vec{k} \in \rar{S}_{j,j+M(j)-1}$  and $\rar{r}_j = 
\rar{r}_j(\vec{k}) \leq \log 
L(1-\epsilon)/\rar{E}_j \leq M(j)$ then 
$$ \pr_{\gamma} \left( k_j,\ldots , k_{j+\rar{r}_j-1} \right) \geq 
\pr_{\gamma}\left( k_j,\ldots ,k_{j+M(j)-1} \right) \geq
\exp(-M(j) E_{\gamma}(1+\epsilon))
= L^{-1 + \epsilon^{2}}.$$
Since these finite strings are distinct, the 
number of strings in $J_i$ such
that these two conditions hold is less than $L^{1-\epsilon^{2}}$. 
Summing over
intervals $J_i$, the total number of such strings is less than
$L(L^{-\epsilon^{2}} N)$. 
Hence if $L \geq C_1(\eta, \epsilon) = (N(\epsilon)/\eta)^{1/\epsilon^2}$
then $L^{-\epsilon^2} N \leq \eta$ and the assertion holds. 

If $\zeta/\log L \rightarrow 0$, then since $N$ grows linearly with $\zeta$,
we know that $L^{-\epsilon^2} N$ still tends to zero as required.
\qed \end{proof}

Note that the precise definition of match length $\rar{r}_i(\vec{k})$ 
doesn't matter, and that
this analysis will go through for a variety of 
related definitions. For
example, the original Lempel--Ziv parsing 
algorithm (see Introduction), or one-sided definitions of match
lengths can be analysed in the same way. The 
key observation is 
that $\vec{k}_i\left(\rar{r}_i\right)$ are 
distinct strings. We deal with these
issues in Section \ref{sec:lz}.

Next, we show that most of the time, we are in the typical set
$\rar{S}_{i,M(i)}$, using a series of 
applications of Chebyshev's inequality.
\begin{lemma} \label{lem:vdec}
Suppose $\zeta$ is fixed, or that $\theta$ satisfies Assumption \ref{ass:theta}
and $\zeta$ satisfies Assumption \ref{ass:zeta}.
For any $\eta>0$, ${\cal P}^{\times L}$-a.s.:
$$ \frac{1}{L} \sum_{i=\eta L+1}^{ \zeta L}
I(\rar{\vec{K}} \notin \rar{S}_{i,M(i)}) \geq \eta,$$
for only finitely many values of $L$. 
\end{lemma}
\begin{proof} We require
\begin{equation} \label{eq:nubd}
\frac{1}{\zeta L}
\left( \sum_{i=\eta L+ 1}^{\zeta L} \nu_i^L \right)
\leq \max_i \nu_i^L \rightarrow 0, \end{equation}
 where 
$\nu_i^L: =  {\cal P}^L(\rar{\vec{K}} \notin \rar{S}_{i,M(i)}).$
The key is a uniform bound on the 4th moment $\ep^L( \rar{Y}_j)^4$ of 
$\rar{Y}_j = \theta(j/L)
(K_i - \ep^L K_i)$.
Note that for $X$ a geometric variable with parameter $q$:
$ \ep (X- \ep X)^4 = q(q^2 + 7q +1)/(1-q)^4 \leq 9/(1-q)^4,$
so that writing $\overline{\theta}$ for $\inf_{x > \eta} \theta(x)$:
$$ 
\ep^L \left( \rar{Y}_j \right)^4 
\leq 9 \left( \frac{\beta (\mu + \theta(j/L))}{1 - e^{-\beta (\mu + \theta(j/L)
)}} \right)^4
\leq 9 \left( \frac{\beta (\mu + \theta(j/L))}{1 - 
e^{-\beta (\mu +\overline{\theta})}} 
\right)^4 = C \left( \mu + \theta(j/L) \right)^4.$$
Hence for any set $S$:
\begin{eqnarray*}
 \ep^L\left(\sum_{j\in S} \left|\rar{Y}_j\right|\right)^4
& \leq & \sum_{j\in S}\ep^L\left(\rar{Y}_j\right)^4 +
3\sum_{j,k\in S,j \neq k}\ep^L\left(\rar{Y}_j\right)^2 \\
& \leq & 3C \max_{j \in S} \left( \mu + \theta(j/L) \right)^4 (\#S)^2.
\end{eqnarray*}
By Chebyshev, for any $i$, 
\begin{eqnarray*}
\nu_i^L
& \leq &  \frac{\ep \left( \sum_{j=i}^{i+M(i)-1} \rar{Y}_j \right)^4}
{(M(i) \theta(i/L) \epsilon')^4} 
\leq  \frac{3C}{\epsilon^{'4}} 
\max_{j \in (i,i+M(i)-1)} \left( \frac{\mu + \theta(j/L)}{\theta(i/L)}
 \right)^4 \frac{1}{M(i)^2},
\end{eqnarray*}
which is less than $K/M(i)^2$, under Assumption \ref{ass:theta}.
So $\max_i \nu_i^L \leq K/(\min_i M(i))^2$, and since $\min_i M(i) =
\log L(1-\epsilon)/\zeta$, if $\zeta /\log L \rightarrow 0$, then
Equation (\ref{eq:nubd}) holds.

Now $Z^L_i = I(\rar{\vec{K}} \notin \rar{S}_{i,M(i)}) - \nu_i$
is a variable with mean 0, variance $\leq \nu_i$ 
and $Z^L_i$ and $Z^L_j$ are independent, if $|i-j| \geq M(i)$.
Hence $\var \left( \sum_{i=\eta L+1}^{\zeta L} Z^L_i \right) 
\leq \sum_{i=\eta L+1}^{\zeta L} \nu^L_i M(i) \leq K \zeta L/(\min_i M(i))
= K \zeta^2 L/((1-\epsilon) \log L)$.

Overall, then, we deduce that for large enough $L$:
\begin{eqnarray*}
{\cal P}^L \left( \frac{1}{\zeta L} 
\sum_{i=\eta L+1}^{\zeta L} I(\rar{\vec{K}} \notin 
\rar{S}_{i,M(i)}) \geq \eta \right) 
& \leq  &  \frac{ \var \left( \sum_{i=\eta L+1}^{\zeta L} Z^L_i \right)}
{(\zeta L)^2 \left(\eta  - \sum_i \nu_i/\zeta L \right)^2} 
 \leq    \frac{4K}{\eta^2 L\log L},
\end{eqnarray*}
which is summable in $L$. \qed \end{proof}
The proof of Lemma \ref{lem:mainlow}
is now complete.
\section{Proof of upper bounds} \label{sec:uppbd}
We establish the 
upper bound in Lemma \ref{lem:mainupp},
by proving a related result about return times.
\begin{definition}
The return time $T_{n,i}(\vec{k})$ 
is how long you have to wait until the substring
$\vec{k}_i(n)$ is repeated in $\vec{k}$:
$ T_{n,i}(\vec{k}) = \inf \{ j \geq 1: \ku{i+j}{n} = \ku{i}{n} \},$
and $\rev{T}_{n,i}(\vec{k})$ is a time-reversed version:
$ \rev{T}_{n,i}(\vec{k}) = \inf \{ j \geq 1: \ku{i-j}{n} = \ku{i}{n} \}.$
\end{definition}
Theorem 1 of \cite{ornstein} shows that 

{\sl For a stationary ergodic probability measure on ${\cal K}$ 
with entropy $h$, for any} $i$:
$$ \lim_{n \rightarrow \infty} \frac{{\log T}_{n,i}(\vec{K})}{n} = h
\;\mbox{ a.s.}$$ 
%
We need a version of this result for distributions ${\cal P}^L$.
In the limit we are close to the IID
case, so we lose little in comparison with that case.

Wyner and Ziv \cite{wyner} 
were the first to exploit the 
dual relationship between waiting times $T_{n,i}$ and match lengths 
$\rar{R}_u$. We shall follow Shields' approach, \cite{shields}, 
modified subsequently in \cite{quas} and \cite{shields2} (to remove
a confusion in \cite{shields} in the way in which return times are defined 
-- whether `overlapping matches', when $T_{n,i} \leq n$, are counted). 

A useful element introduced in \cite{quas} is a truncation argument
needed to cover the case of geometric random variables (the 
analysis in \cite{shields} only holds for finite alphabet processes). 
\cite{quas} introduces a truncation operation $\tau_m$ where $\tau_m(x)
=\min(x,m)$ and $\tau_m(\vec{x}) = \left(\tau_m(x_i),\,i\in \Z\right)$.
Denote the match lengths and entropies of the truncated process by
$\rart{r}_i(\vec{k})$, $\rart{R}_i(\vec{K})$ and $\rart{E}_i$.
First note that
$\rar{r}_i(\vec{k})\leq\rar{r}_i(\tau_m(\vec{k}))=$ 
$\rart{r}_i(\vec{k})$.
Secondly, since $\rart{E}_i/\rar{E}_i =
1 - \exp(- \beta (\mu + \theta(i/L)m)$, we can ensure that
$\rar{E}_i<\rart{E}_i(1- \epsilon/2)(1+ \epsilon)$ for all $i$. 
Hence, we need only prove that:

\begin{lemma} \label{lem:mainupper}
For fixed $\theta$, $\eta$, $\epsilon$, then for each string $\vec{k}$
defining:
$$ \rar{U}(\vec{k}) = \left\{ i > \eta L: \rart{r}_i(\vec{k}) > 1 + 
\frac{{\log L}}{\rart{E}_i (1- \epsilon/2)} \right\} $$
then $\;\limsup_{L\to\infty} \# \rar{U}(\vec{K})/L \leq 
\eta$, ${\cal P}^{\times L}$-a.s. \end{lemma}
\begin{proof}
We mirror the duality argument (cf Lemma 3 of \cite{shields}
and Appendix of \cite{shields2}). 
Define for $N=1,2,\ldots$ the forward count:
$$ \rar{F}_N(\vec{k}) 
=\left\{i:\frac{\log T_{n,i}(\vec{k})}{n}<\rart{E}_i(1-\epsilon/2) 
\mbox{ for some $n \geq N$} \right\},$$
and backwards count:
$$ \rar{B}_N(\vec{k}) = 
\left\{i:\frac{\log \rev{T}_{n,i}(\vec{k})}{n} < \rart{E}_i(1-\epsilon/2) 
\mbox{ for some $n \geq N$} \right\},$$
Now, if $i \in \rar{U}(\vec{k})$, then there 
exists $j \neq i$ such that $\vec{k}_i(s)
= \vec{k}_j(s)$, where $s=\rart{r}_i(\vec{k}) -1$, so either
\begin{enumerate}
\item{If $i < j$, then $T_{n,i}(\vec{k})\leq L$, so that 
$\log T_{n,i}(\vec{k})/n\leq\log L/n<\rart{E}_i(1-\epsilon/2)$}
\item{If $j < i$, then $\rev{T}_{n,i}(\vec{k}) \leq L$, 
so that $\log \rev{T}_{n,i}(\vec{k})/n 
\leq\log L/n <\rart{E}_i(1-\epsilon/2)$}
\end{enumerate}

Hence if $i \in \rar{U}(\vec{k})$, then 
$i$ is in $\rar{F}_N(\vec{k})$ or $\rar{B}_N(\vec{k})$ 
for some $N$. So, using the finiteness of the alphabet, if we can show
that $\rar{F}_N(\vec{K})$ and $\rar{B}_N(\vec{K})$ are of low density
(that is $|\rar{F}_N(\vec{K})|/L$ and $|\rar{B}_N(\vec{K})|/L$ are
$\leq \eta$ eventually, ${\cal P}^{\times L}$-a.s.)
then so must $\rar{U}(\vec{K})$ be, and the result follows.

First, we show that the number of 
overlapping matches is small. We mirror \cite{quas}
and define
${\cal A}=\left\{\vec{k}: \vec{k}_1(s)=\vec{k}_{s+1}(s)\;
\mbox{ for infinitely many $s\geq 1$} \right\}.$
We will show that this set has measure 0, by defining
${\cal B}_m=\left\{\vec{k}: \vec{k}_1(m)=\vec{k}_{m+1}(m)\right\},$
so that ${\cal A}=\bigcap_l\bigcup_{m \geq l}{\cal B}_m$. 
Following \cite{quas}, for each $m$ and $\vec{w} \in \Z_+^m$, write
${\cal W}(\vec{w})$ for the set of strings which 
begin with word (i.e., have $\vec{k}_1(m)=\vec{w}$),
${\cal W}(\vec{ww})$ for the
strings which begin with $\vec{w}$ repeated twice. 
Now for $\delta >0$ consider a $\delta$-representative set
${\cal V} = \{ \vec{w}: - \log {\cal P}^L(\vec{w}) \geq \sum_{j=1}^m
\rar{E}_j - \delta \}.$
On set ${\cal V}_{\delta}$, since the entropy is bounded below:
\begin{eqnarray*}
{\cal P}^L ({\cal B}_m \cap{\cal V}_{\delta}) & = & \sum_{\vec{w}} 
{\cal P}^L \left( {\cal W}(\vec{ww}) \right) \\
& \leq & \left( \sum_{\vec{w}}{\cal P}^L \left( {\cal W}(\vec{w})
\right) \right) 
\exp \left( -\sum_{j=m+1}^{2m} \rar{E}_j +\delta \right) \\
& = & \exp \left( -\sum_{j=m+1}^{2m} \rar{E}_j +\delta \right),
\end{eqnarray*}
which is summable in $m$. So a Borel-Cantelli argument 
establishes the result.

Next, to bound $\rar{F}_N(\vec{K})$, 
we consider a word $\vec{x} = (x_1,\ldots ,x_{n})$ which lies in a 
$\delta$-representative set of 
the $\vec{k}_i(n)$'s, that is for some $\delta >0$:
\begin{equation} \label{eq:typset}
\sum_{l=1}^n | x_l-\ep^LK_{i-1+l}| < \delta/\beta. \end{equation} 
By direct calculation, we can bound from above the probability that 
$\vec{x}$ turns up later, that is for $j > i$:
$${\cal P}^L\left(\vec{K}_j(n) =\vec{x} \right) 
\leq{\cal P}^L\left(\vec{K}_i(n)=\vec{x} \right)
\exp(\theta^* \delta).$$

Hence for any integer $t$, if Equation (\ref{eq:typset}) holds:
\begin{eqnarray*}
{\cal P}^L \left( n +1 \leq T_{n,i}(\vec{K})\leq t\left|
\vec{K}_i(n) =\vec{x} \right) \right.
& = & \sum_{m=i+n+1}^{i+t}{\cal P}^L \left(\vec{K}_m(n) 
=\vec{x} \left| \vec{K}_i(n)=\vec{x} \right) \right. \\
& \leq & t{\cal P}^L\left(\vec{K}_i(n)=\vec{x} \right)\exp(\theta^* \delta) \\
& \leq & t \exp(-n \rar{E}_i +2\theta^* \delta).
\end{eqnarray*}
Then with $t= \exp(n(\rar{E}_i-\epsilon))$, we need to pick $\delta$ 
growing slowly enough that $\theta^* \delta/n$ tends to zero -- say
$\delta = n^{7/8}$ (if
$\zeta_L$ is growing more slowly than 
$\log L$, we can still choose appropriate $\delta$).
Consider overlapping and non-overlapping matches separately:
$${\cal P}^L\left(\left.\frac{\log T_{n,i}(\rar{\vec{K}})}{n} \leq 
\rar{E}_i-\epsilon\right|\vec{K}_i(n)=\vec{x} \right)
\leq {\cal P}^L \left( \bigcup_{m \geq n/2} {\cal{B}}_m \right) 
+ \exp \left( - \frac{n \epsilon}{2} \right),$$
for $n$ sufficiently large. 
As $n \rightarrow \infty$, the probability that Equation (\ref{eq:typset})
holds tends to 1. 
We can bound the backward set $\rar{B}_N(\vec{K})$ similarly. 
\qed \end{proof}
We can now prove the uniform upper bound 
in a more straightforward fashion

\begin{pfof}{of Lemma \ref{lem:unifupp}} 
Since for
any $j$, $\max_i {\cal P}^L(K_j = i) = {\cal P}^L(K_j = 0) =
1 - \exp(- \beta (\mu + \theta(j/L))) \leq  
1 - \exp(-\beta (\mu + \theta^*))$, then for any $N$:
\begin{eqnarray*}
{\cal P}^L\left(\rar{R}_i(\vec{K}) \geq N \right) &= &
{\cal P}^L\left(\vec{K}_j(N)=\vec{K}_i(N), 
\mbox{ for some } j\in\{1,\ldots ,L\},j\neq i \right)  \\
& \leq & \sum_{j=1\neq i}^L {\cal P}^L\left(\vec{K}_j(N)
=\vec{K}_i(N)\right) \leq L(1 - \exp(-\beta (\mu + \theta^*)))^N. 
\end{eqnarray*}
So ${\cal P}^L(\max_i \rar{R}_i \geq N) \leq L^2 (1 - \exp(-\beta (\mu 
+ \theta^*)))^N$.
Taking $c > -3/\log (1 - \exp(-\beta (\mu + \theta^*)))$, and $N = c 
\log L$, the result follows. 

Again, if $\zeta_L/\log L \rightarrow 0$, the same bounds will work:
since we need to make more comparisons, replace $L^2$ by
$(L \zeta_L)^2$, and the logarithmic term is dominated by the polynomial.
\qed \end{pfof}
\section{Fermions} \label{sec:fermdir}
We can use the same techniques to consider the 
alternative model of two-point random variables
(still in one dimension).
We make the following observations, which ensure 
that the above proofs will carry through.
\begin{enumerate}
\item{We adapt the proof of Lemma \ref{lem:propbd}, 
introducing, for $0<p<1$:
$$D_{p,M}=\left\{\vec{x}_1(M):\sum_{i=1}^Mx_i\leq M \left( p + 
\frac{\epsilon E_p}{\log(1/p-1)} \right) \right\}. $$
Here $E_p$ stands for the entropy $-p\log\,p-
(1-p)\log\,(1-p)$.
Again, it is true that for $\vec{x}_1(M)\in D_{p,M}$,
$\pr_p(\vec{u}) \geq \exp(-ME_p(1+\epsilon))$,
where $\pr_p$ is the Bernoulli measure on ${\cal K}_-
= {\K}_-^ \Z$, with $\pr_p (K_i=0)
= 1-p$, $\pr_p (K_i=1) = p$.
The assertion of Lemma    \ref{lem:propbd}   then follows in the 
same way as before.}
\item{For a string $\vec{k} = (k_i, i \in \Z) \in {\cal K}$, we define 
$$\rar{y}_i(k_i) = -\log{\cal P}^L(K_i=k_i)-\rar{E}_i
=  (k_i - \ep^LK_i)\log \left(1/{\cal P}^L(K_i=1) -1\right),$$ 
and for the random string $\vec{K}$, define $\rar{Y}_i(K_i)$
in the same fashion.}
\item{For random variable $K$ taking values $0$ with 
probability $1-p$ and $1$ with probability $p$, if $Y(k) =
-\log \pr(K=k) - E_p$ then
$$\ep \left( Y(k)\right)^4=p(1-4p+6p^2-3p^3)\log\,(1/p-1)^4.$$
Since for $p \in [0,1]$: $1-4p+6p^2-3p^3\leq 1$, and making the
substitution $y=\log(1/p-1)$, for $p \leq 1/2$ implies 
$p\log(1/p -1)^4 = y^4/(1+e^y) \leq 24$. By 
symmetry, the same result holds for $p > 1/2$.
Hence the proof of Lemma \ref{lem:vdec} goes through.}
\item{Since we now deal with finite alphabets only, 
the proof of Lemma
\ref{lem:mainupp} simplifies. We don't need the truncation
argument previously described, and our observations 
about representative sets will go through as before.}
\item{The upper bound in Lemma \ref{lem:unifupp} 
is proved in the same way, since a uniform bound
$\max_{i,j} 
\pr( \rar{X}_j = i) \leq \max( 1/(1+e^{-\beta (\mu + \theta^*))}, 
  1/(1+e^{\beta \mu}))$ 
holds.}
\end{enumerate}
\section{Adaptions to the higher-dimensional case} \label{sec:highdim}
As in \cite{quas}, the generalization to higher dimensions goes 
through in a rather straightforward fashion.
\begin{enumerate}
\item{The proof of Lemma \ref{lem:propbd} carries
through; we still divide the
larger region  into sets $J_i=\{ \vec{u}= (u_1,\ldots u_d)\in
 \Z^d$: $\nm{\vec{u}/L} \in (u_i, u_{i+1})$ 
on which the variables are nearly IID. In general we 
need to replace $M$ by $M^d$, so for example:
$$D_{\gamma, M}=\left\{\vec{x}_{\vec{1}}(M): 
\sum_{i\in \bx{1}{M}}x_i\leq M^d \left( 
\frac{1}{e^{\gamma} -1} + \epsilon 
\frac{E_{\gamma}}{\gamma} \right)\right\}. $$
We introduce $M(i) = (d \log L (1-\epsilon)/
E_{u_{i+1}})^{1/d}$.}
\item{The proof of Lemma \ref{lem:vdec} 
goes through as before, since the 
uniform bound on the 4th moment of $\rar{Y}_{\vec{j}}$ 
still holds.}
\item{We can extend the definition of waiting 
time required in Section \ref{sec:uppbd}. 
Writing $\vec{v} = (v_1,\ldots ,v_d) \geq 0$ 
to mean that $v_l \geq 0$, $1 \leq l\leq d$,
and with $|\vec{v}|^+ = \max v_l$:
$$ T_{n,\vec{u}}(\vec{k})=\inf \{|\vec{v}|^+: \vec{v} \geq 0,\vec{k}_{\vec{u}+
\vec{v}}(n) =\vec{k}_{\vec{u}}(n) \}.$$}
\item{ The upper bound in
Lemma \ref{lem:unifupp} is proved in the same
way, since a uniform bound on $\max \left[ {\cal P}^L(K_{\vec{u}} = j),
\vec{u} \in  \Z^d, j \geq 1 \right]$ holds.}
\end{enumerate}
\section{Lempel-Ziv parsing} \label{sec:lz}
Now  we establish the 
Lempel-Ziv parsing algorithm for one-dimensional 
free quantum systems. We use the notation from Sections \ref{sec:lowbd}--
\ref{sec:fermdir}.
Recall the algorithm takes a string (or a `message') 
$\vec{k}_1(L)$ and parses it
into words; at each stage, we add a marker, `;', so that 
the parsed block is the shortest word not already seen.
\begin{definition}[Lempel-Ziv parsing]
We parse the string $\vec{k}_1(L)=(k_1,\ldots $,\\ $k_L)$ 
into words:
$$\vec{k}_1(L)=\{ \vec{k}_{t(1)}(l(1)) ; \vec{k}_{t(2)}(l(2)) ; 
\ldots ; \vec{k}_{t(c)}(l(c)) ; \vec{k}_{t(c)+1}(r) \},$$
according to the rule: $t(1)=1$, $t(i+1)=t(i)+l(i)$, 
$$l(i+1)=\min\,\{ m \geq 1: \vec{k}_{t(i)}(m)
\notin \{\vec{k}_{t(1)}(l(1), \ldots ,\vec{k}_{t(i)}(l(i))\} \},$$
where $\vec{k}_{t(c)+1}(r)$ is the remaining word, $r=L-t(c)-1$
and $c+1$ ($=c(\vec{k},L)+1$) the total number of parsed words.
\end{definition}

As was noted in the Introduction,
this parsing rule is associated with a data-compression algorithm
which is asymptotically efficient (achieves the upper bound provided by
entropy) for ergodic processes. The algorithm relies on the fact that for
each word $\vec{k}_{t(i)}(l(i))$, we can describe it by first giving the 
point in the string between 1 and $t(i) \leq L$ where block $
\vec{k}_{t(i)}(l(i)-1)$ previously occurs, and then 
by giving the extra symbol which is different. 
Thus we require $\log L +1$ symbols to specify each parsed word in 
$\vec{k}_1(L)$ and the total length of the compressed message will be:
$ c(\vec{k},L) (\log L +1)$, cf Shields \cite{shields3}, Chapter 11.
\begin{theorem} \label{thm:lzopt}
For the one-dimensional quantum free ensemble, 
for all $\zeta >0$,
\begin{equation} \label{eq:lzopt}
\lim_{L\to\infty} \frac{c(\vec{K}_1(\zeta L))\log L}{L} =
h_{\pm}^{[0,\zeta ]}, \mbox{  ${\cal P}_\pm^{\times L}$-a.s.,}\end{equation}
Under Assumptions \ref{ass:theta} and \ref{ass:zeta}:
$$\lim_{L\to\infty} 
\frac{c(\vec{K}_1(\zeta L))\log L}{L} =
h_\pm, \mbox{  ${\cal P}_\pm^{\times L}$-a.s.} $$
\end{theorem}
\begin{proof}
We know that the
RHS is $\lim_{L\to\infty} \sum_{i=1}^L \rar{E}_{i}/L$ 
which represents the data compression limit. That is, Shannon's Noiseless
Coding Theorem (see for example Theorem 5.3.1 of \cite{cover}) states that
the expected length of any decipherable code for a random variable $X$ is
greater than or equal to the entropy of $X$.

Therefore, to prove Equation (\ref{eq:lzopt}), 
it remains to establish the upper bound $ \limsup_{L\to\infty} 
c(\vec{K}_1(\zeta L)) \log L/L  \leq h^{[0,\zeta ]}$.
We prove this using analysis similar to that of 
Section \ref{sec:lowbd}.
As before, the proof goes in the same way 
for all values of $\zeta$, so we fix $\zeta =1$.
Once again, we split the interval $[0,1]$ into 
subintervals $J_i =(u_i, u_{i+1})$, 
and for each $i$, write $\vec{k}_{J_i}$ for $(k_{Lu_i},
\ldots ,k_{Lu_{i+1}-1})$ and set 
$G_i=\{t(j): Lu_i\leq t(j)\leq L u_{i+1}\}$ 
(the start-points of words which lie within the sub-interval). We also
put $N_i = \{ r \in G_i: \sum_{j=r}^{r+l(r)-1} \rar{E}_j 
\leq \log L(1- \epsilon) \}$.

In the spirit of Lemma \ref{lem:propbd}, we first observe that the
cardinality $|N_i|
\leq L^{1-\epsilon^2}$, since again, these parsed words are short distinct
strings, in the typical set. Then, considering the entropy present in these
parsed words, we deduce that:
$$
\sum_{j \in J_i} \rar{E}{L}_j =  \sum_{r \in G_i} 
\left( \sum_{s=r}^{r+l(r)-1} \rar{E}{L}_s \right) 
\geq  \log L (1-\epsilon) (|G_i| - N_i). $$
On rearranging we deduce that
$$ \limsup_{L\to\infty} \frac{|G_i|
\log L}{L} \leq \frac{\sum_{j \in J_i} \rar{E}{L}_j}{L} + \epsilon.$$
The theorem follows by summing over intervals $J_i$, since $\sum
|G_i| = c(\vec{k},L)$. 

We can deal with the case of $\zeta/\log L \rightarrow 0$ as before.
\qed \end{proof}

\begin{acknowledgement}
YS thanks Prof. V. Zagrebnov (CPT, Marseille-Luminy) for fruitful discussions,
and IHES, Bures-sur-Yvette, for hospitality during visits in February, June
and September 2001. OJ is a fellow of Christ's College Cambridge. Both 
authors are part of the Cambridge-MIT Institute collaboration `Quantum
Information Theory and Technology'.
\end{acknowledgement}


\begin{thebibliography}{C-N-T}

\bibitem[B-R]{br2}
Bratteli, O. and Robinson, D.W.: {\em Operator algebras and quantum
statistical mechanics 2}. Berlin: Springer, 1997. 

\bibitem[C-N-T]{connes}
Dynamical entropy of $C^*$ algebras and von Neumann algebras.
Commun. Math. Phys. \textbf{112}, 691--719 (1987)

\bibitem[C-T]{cover}
Cover, T.M. and Thomas, J.A.:
{\em Elements of Information Theory}.
New York: John Wiley, 1991.

\bibitem[G]{grassberger}
Grassberger, P.:
Estimating the information content of symbol sequences and efficient
  codes.
IEEE Trans. Inform. Theory. \textbf{35}, 669--675 (1989)

\bibitem[J-K-P]{jkp}
Julsgaard, B., Kozhekin, A. and Potzik, E.S.:
Experimental long-lived entanglement of two macroscopic objects.
Nature. \textbf{413}, 400--403.

\bibitem[K-S]{kontoyiannis}
Kontoyiannis, I. and Suhov, Y.
Prefixes and the entropy rate for long-range sources.
In Kelly, F.P. (ed.) {\em Probability, Statistics and Optimisation}, pp.
89--98. New York: John Wiley, 1993.

\bibitem[L]{lieven}
Lieven, M.K.V., Steffen, M., Breyta, G., Yannoni, C.S., Sherwood, M.H. and
Chuang, I.L.: 
Experimental Realisation of Shor's quantum factoring algorithm using
nuclear magnetic resonance.
Nature. \textbf{414}, 883--887.

\bibitem[N-C]{nielsen}
Nielsen, M.A. and Chuang, I.L.:
{\em Quantum Computation and Quantum Information}.
Cambridge: CUP, 2000.

\bibitem[O-W]{ornstein}
Ornstein, D.S. and Weiss, B.:
Entropy and data compression schemes.
IEEE Trans. Inform. Theory. \textbf{39}, 78--83 (1993)

\bibitem[Q]{quas}
Quas, A.N.:
An entropy estimator for a class of infinite processes.
Theory Probab. Appl. \textbf{43}, 496--507 (1999)

\bibitem[S]{sackett}
Sackett, C.A., Klepinski, D., King, B.E., Langer, C., Meyer, V., Myatt, C.J.,
Rowe, M., Turchette, Q.A., Itano, W.M., Wineland, D.J. and Munroe, C.:
Experimental entanglement of four particles.
Nature \textbf{404}, 256--259 (2000)

\bibitem[Sh1]{shields}
Shields, P.C.:
Entropy and prefixes.
Ann. Probab. \textbf{20}, 403--409 (1992)

\bibitem[Sh2]{shields2}
Shields, P.C.:
String matching bounds via coding.
Ann. Probab. \textbf{25}, 329--336 (1997)

\bibitem[Sh3]{shields3}
Shields, P.C.:
{\em The ergodic theory of discrete sample paths}.
Providence, RI: American Mathematical Society, 1996.

\bibitem[W-Z]{wyner}
Wyner, A.D. and Ziv, J.:
Some asymptotic properties of the entropy of a stationary ergodic
  data source with applications to data compression.
IEEE Trans. Inform. Theory. \textbf{35}, 1250--1258 (1989)

\bibitem[Z-L]{lempel}
Ziv, J. and Lempel, A.:
A universal algorithm for sequential data compression.
IEEE Trans. Inform. Theory, \textbf{23}, 337--343 (1977)



\end{thebibliography}
\end{document}